\documentclass[prd,preprint,tightenlines,floatfix,showpacs,preprintnumbers,nofootinbib,eqsecnum,superscriptaddress]{revtex4}

 \usepackage[dvips,final]{graphicx}
  \usepackage{amssymb}
   \usepackage{amsmath}
    \usepackage{amsfonts}
     \usepackage{epsfig}
      \usepackage{bm}

\usepackage[section]{placeins}

\usepackage{multirow}
\usepackage{ctable}
\usepackage{booktabs}
\usepackage{array}
\usepackage{tabularx}
\usepackage{xcolor}
\usepackage{pstricks}

\def\Pom{{\bf I\!P}}

\def\lsim{\mathrel{\rlap{\lower4pt\hbox{\hskip1pt$\sim$}}
    \raise1pt\hbox{$<$}}}         
\def\gsim{\mathrel{\rlap{\lower4pt\hbox{\hskip1pt$\sim$}}
    \raise1pt\hbox{$>$}}}         

\begin{document}
\title{Photoproduction of $J/\psi$ mesons \\
in peripheral and semi-central heavy ion collisions}

\author{Mariola K{\l}usek-Gawenda}
\email{mariola.klusek@ifj.edu.pl} \affiliation{Institute of Nuclear
Physics PAN, PL-31-342 Cracow,
Poland}

\author{Antoni Szczurek}
\email{antoni.szczurek@ifj.edu.pl} 
\altaffiliation{also at University of Rzesz\'ow, PL-35-959 Rzesz\'ow, Poland}
\affiliation{Institute of Nuclear
Physics PAN, PL-31-342 Cracow,
Poland}


\date{\today}

\begin{abstract}
We calculate total and differential cross sections
for $J/\psi$ photoproduction in ultrarelativistic lead-lead collisions
at the LHC energy $\sqrt{s_{NN}}=2.76$ TeV.
In the present approach we use a simple model based on vector dominance
picture and multiple scattering of the hadronic ($c \bar c$) state 
in a cold nucleus as an example.
In our analysis we use both the classical mechanics and quantum (Glauber)
formulae for calculating $\sigma_{tot}(J/\psi Pb)$ which is a building
block of our model.
We compare our UPC results with ALICE and CMS data.
For semi-central collisions ($b<R_A+R_B$)
a modification of the photon flux seems necessary. 
We discuss different physics motivated approximations.
We try to estimate the cross sections for different centrality bins
and for $J/\psi$ mesons emitted in forward rapidity range ($2.5<y<4$)
corresponding to recent ALICE experimental results. 
Reasonable results are obtained but open questions are discussed.

\end{abstract}


\maketitle

\section{Introduction}
\label{introduction}

Last years the ALICE Collaboration studied production of 
$J/\psi$ mesons mostly in central $^{208}$Pb--$^{208}$Pb collisions 
(see e.g. \cite{ALICE_midrapidity}) at $\sqrt{s_{NN}}$ = 2.76 TeV i.e. 
at the highest available nucleus-nucleus center-of-mass energies. 
At these high energies different theoretical mechanisms come into game 
(see e.g. \cite{Rapp,Andronic,Zhao_Rapp,Zhou,Blaizot} and references therein). 
For instance in Ref. \cite{Andronic} the authors
emphasized thermal aspect of charmonia production. 
In order to describe detailed distributions in transverse momentum, 
rapidity or multiplicity several dynamical mechanisms have to be 
taken into account. In contrast to early expectations \cite{Satz}
there are mechanisms which lead to both suppression
and regeneration of $J/\psi$ quarkonia in cold and hot medium.
Experimental studies discuss usually
so-called nuclear modification factor as a function of
multiplicity, transverse momentum and rapidity of $J/\psi$
and smooth dependences have been observed 
(a nice summary was presented e.g. in \cite{Munzinger_Trento}).

In last years there was also an interest in 
calculating cross sections for exclusive production
of $J/\psi$ mesons in ultrarelativistic heavy ion collisions
\cite{RSZ_2012,KN_1999,AB_2012,GM_2011,CSSz_2012,DGM_2013}. 
The main reason for these investigations
was a better understanding of interaction of a small size 
($c \bar c$) perturbative dipole
with a nucleus or more explicitly with the nuclear gluons
incorporated in different ways in different models.
In general, this may be related to gluon saturation phenomenon 
which in the case of nucleus may be more efficient than for 
the case of nucleon due to a coherent action of many nucleons 
(larger gluon occupation).
All the previous investigations were related to purely
ultraperipheral collisions (UPC), i.e. when the nuclei stay intact
\footnote{An emission of extra neutrons caused by additional purely
electromagnetic interactions could be included too.}.
In theoretical calculations UPC means in practice $b > R_A + R_B$
(sum of nuclear radii).
All the calculations must be therefore performed in the impact parameter
space to include this condition.

Can the photoproduction process (photon emission, its fluctuation 
into $c \bar c$ and subsequent rescattering of the dipole 
(or hadron) in the nucleus) be active also for less peripheral collisions 
i.e. when nuclei collide and break apart producing, 
at these high energies, quark-gluon plasma?
A recent analysis of the ALICE Collaboration \cite{ALICE_forward} 
presents differential studies of inclusive production of 
$J/\psi$ in Pb-Pb collisions exclusively at forward rapidities.
At low transverse momenta and small multiplicity they observe 
an enhancement of $R_{AA} >$ 1. According to our knowledge this
enhancement was not explained so far in the literature.
We wish to address this issue in the following paper.
We will try to discuss the issue in the equivalent photon approximation (EPA).
In the present first estimate of the cross section we 
will discuss several approximations how to calculate relevant photon 
fluxes to include approximately the physical conditions relevant
in the photoproduction of $J/\psi$.
This is very important for the peripheral and semi-central
collisions.
For simplicity in the following we will use a simple vector dominance 
model (VDM) combined with multiple scattering
of the hadronic photon fluctuation for calculating 
the $\gamma A$ cross section which, 
as will be shown in this paper, describes relatively well the cross section
for production of $J/\psi$ mesons in UPCs.

A very recent analysis of the ALICE Collaboration \cite{ALICE_presentation}
confirms the presence of the enhancement at very small $J/\psi$
transverse momentum and tries to extract the new not fully understood
contribution for different centrality bins.
We shall try to describe the ALICE preliminary data
\cite{ALICE_presentation} assuming the mentioned above
photoproduction mechanism.

\section{Description of our simple model}
\label{nuc_photo}

Nuclear photoproduction of a single vector meson $V$
can be understood as a photon fluctuation into hadronic 
(virtual $V$ meson or quark-antiquark pair) 
component and its subsequent propagation through the second (cold)
nucleus and a transformation (fragmentation) into the on-shell $V$ meson.

\begin{figure}[h!]
\centering
\includegraphics[scale=0.4]{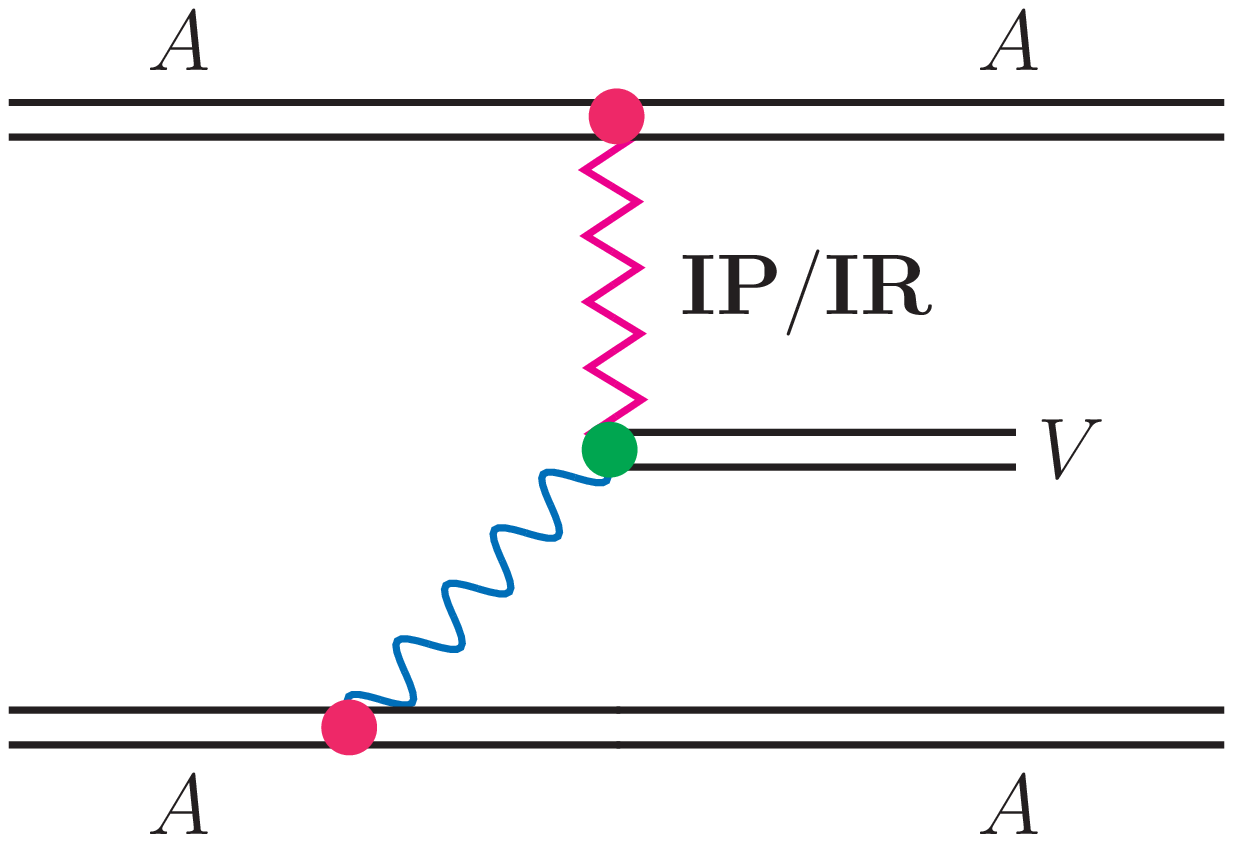}
\includegraphics[scale=0.4]{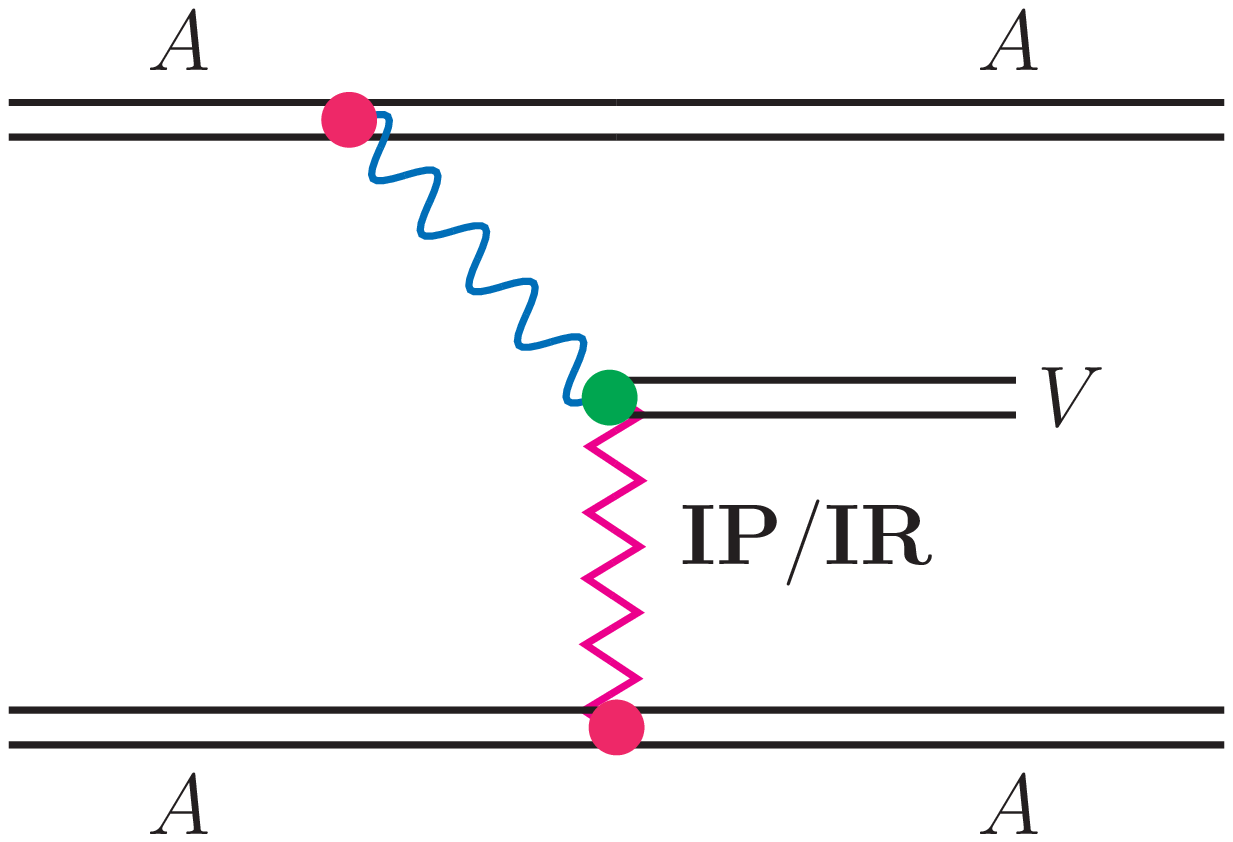}
\caption{Schematic diagrams for the single vector meson production 
by photoproduction photon-Pomeron (left) or Pomeron-photon (right) fusion.
Here the word Pomeron and its symbol is an abbreviation
for multiple diffractive scattering of the hadronic system
in the nuclear medium.}
\label{fig:photoproduction_single}
\end{figure}

Fig.~\ref{fig:photoproduction_single} illustrates a mechanism of
a single vector meson production in ultraperipheral
ultrarelativistic heavy-ion collisions. 
The cross section for this mechanism is usually written
differentially in the impact parameter $b$ and in the vector 
meson rapidity~$y$
\begin{equation}
\frac{\mathrm{d} \sigma_{A_1A_2 \to A_1A_2V}}{\mathrm{d}^2 b \mathrm{d}y} = 
\frac{\mathrm{d}P_{\gamma \Pom}\left(b,y\right)}{\mathrm{d}y} + 
\frac{\mathrm{d}P_{\Pom \gamma}\left(b,y\right)}{\mathrm{d}y} \;.
\label{eq:single_vector_meson}
\end{equation}
Above $P_{\gamma \Pom}(y,b)$ or $P_{\Pom \gamma}(y,b)$ is the probability 
density for producing a vector meson $V$ at rapidity $y$ 
for fixed impact parameter $b$ of the heavy-ion collision.
Probability density expresses two-different possibilities
of the production of vector meson shown in 
Fig.~\ref{fig:photoproduction_single}. 
Each of the probabilities is the convolution
of the cross section for $\gamma_1A_2 \to J/\psi A_2$ or $\gamma_2A_1 \to J/\psi A_1$
(the photon emitted from first or second nucleus
($\omega_{1/2} = m_{V}/2 \exp(\pm y)$)) and a corresponding flux of
equivalent photons:
\begin{equation}
\frac{\mathrm{d}P_{1/2}\left(y,b\right)}{\mathrm{d} y} = 
\omega_{1/2} N\left(\omega_{1/2},b\right) \sigma_{\gamma A_{2/1} \to V A_{2/1}}(W_{\gamma A_{2/1}}) S\left(b\right) \;,
\label{eq:probab_P12}
\end{equation}
where $N\left(\omega_{1/2},b\right)$ is usually 
a function of impact parameter between heavy ions ($b$) 
and not of photon-nucleus impact parameter.
Finally $S(b)$ is an impact parameter dependent survival factor 
which with good precision can be approximated as $S(b) \approx \theta(b-R_A-R_B)$. 

Now we wish to proceed to semi-central collisions
i.e. to the case of $b<R_A+R_B$.
Then the survival factor has to be omitted as we are interested
also in situations when colliding nuclei break apart.
In the present approach we will modify only photon fluxes
and leave the coherent $\gamma A \to J/\psi A$ cross section
unmodified. This seems a crude approximation which should be
reasonable only for rather peripheral collisions $b \approx R_A + R_B$.
We shall try to see how far this approximation can be
extrapolated down to smaller $b$.
The effective photon flux which includes the geometrical aspects 
can be formally expressed through 
the real photon flux of one of the nuclei and effective 
strength for the interaction of the photon with the second nucleus
\begin{equation}
 N^{(1)}\left(\omega_1,b\right) = \int N\left(\omega_1,b_1\right)
\frac{\theta(R_A - b_2)}{\pi R_A^2} \mathrm{d}^2 b_1 
\; ,
\label{eq:d2N/domegadb}
\end{equation}
where ${\bf b_1} = {\bf b} + {\bf b_2}$. 
The extra $\theta(R_A - b_2)$ factor
ensures collision when the photon hits
the nucleus-medium. For the photon flux in the second nucleus
one needs to replace 1$\to$2 (and 2$\to$1).
For large $b \gg R_A+R_B$: $N^{(1)}\left(\omega_1,b\right) \approx N(\omega_1,b)$.
For small impact parameters this approximation is,
however, not sufficient. 
This has some consequences also for ultraperipheral
collisions, which will be discussed somewhat later in this section.

Since it is not completely clear what happens in the region 
of overlapping nuclear densities we suggest another approximation 
which may be considered rather as lower limit. 
In this approximation we integrate the photon flux of 
the first (emitter) nucleus only over this part of the second 
(medium) nucleus which does not collide with the nucleus-emitter
(some extra absorption may be expected in the tube
of overlapping nuclei). 
This may decrease the cross section for more central collisions. 
In particular, for the impact parameter $b=0$ the resulting vector meson 
production cross section will fully disappear by the construction.
In the above approximation the photon flux can be written as: 
\begin{equation}
N^{(2)}\left(\omega_1,b\right) = \int N\left(\omega_1,b_1\right)
\frac{\theta(R_A - b_2) \times \theta(b_1 - R_A)}
{\pi R_A^2} \mathrm{d}^2 b_1  \; .
\label{eq:d2N/domegadb_second}
\end{equation}

We shall use the three different approximations how to calculate
photon fluxes and three different form factors
which are main ingredients of the photon flux.
We shall start with
\begin{equation}
F_{pl}(q) = 1 
\end{equation}
corresponding to point-like charge.
In this approximation the flux of the photons is given by the 
simple formula 
\begin{equation}
\frac{\mathrm{d}^3N(\omega,r)}{\mathrm{d}\omega \mathrm{d}^2r} =
\frac{Z^2\alpha_{em}X^2}{\pi^2\omega r^2} K_1^2(X) \; ,
\end{equation}
where $Z$ is the nuclear charge, $\omega$ is energy of the photon,
$r$ is a distance in the impact parameter space between 
the photon and the emitting nucleus,
$K_1$ is a modified Bessel function 
and $X=r \omega / \gamma$.
The photon flux integrated over all $r$ is approximately
equal to the photon flux in the region of $r> R_A+R_B$. 
This can be calculated analytically:
\begin{equation}
\frac{\mathrm{d}N(\omega)}{\mathrm{d}\omega} = 
\frac{2Z^2 \alpha_{em}}{\pi \omega}
\left( \chi K_0(\chi) K_1(\chi) - \frac{\chi^2}{2} \left[ K_1^2(\chi) - K_0^2(\chi) \right] \right) \;,
\end{equation} 
where $\chi = 2R_A\omega/\gamma$.

In our calculations we use also a form factor which is 
Fourier transform of the charge distribution in nucleus.
Two-parameter Fermi distribution (called equivalently Woods-Saxon distribution)
\begin{equation}
\rho(r) = \frac{\rho_0}{1+\exp\left( \frac{r-c}{a} \right)} \;,
\label{eq:density}
\end{equation}
where the normalization constant $\rho_0$ 
for the lead nuclei equals to $\frac{Z}{A}0.1572$ fm$^{-3}$
and $c=6.62$ fm, $a=0.546$ fm is used in the calculation.
The form factor (called here realistic form factor for brevity)
is calculated then as
\begin{equation}
F_{real}(q^2) = \frac{4\pi}{q} \int \rho(r) \sin(qr) r \mathrm{d}r \;.
\label{eq:F_real}
\end{equation}
In the literature often a monopole form factor is used
\begin{equation}
F_{mon}(q^2) = \frac{\Lambda^2}{\Lambda^2+q^2}
\end{equation}
which simplifies the calculations.
The value of $\Lambda$ can be expressed through
the root mean square electric radius
\begin{equation}
\sqrt{<r^2>} = \sqrt{\frac{6}{\Lambda^2}}
\end{equation}
giving $\Lambda = 88$ MeV for $\sqrt{<r^2>}=5.5$ fm for $^{208}$Pb.
In our calculation we use the following generic formula for
calculating the photon flux (see e.g. Ref. \cite{KGS_1997}) for any
nuclear form factor
%
\begin{equation}
N(\omega,b) = \frac{Z^2 \alpha_{em}}{\pi^2} \left| \int u^2 J_1\left(u\right)
\frac{F\left( \frac{\left( \frac{\omega b}{\gamma} \right)^2+u^2}{b^2} \right)}{\left(\frac{\omega b}{\gamma} \right)^2+u^2} \right|^2 \;.
\label{eq:N_omegab}
\end{equation}

\begin{figure}[!h]
\includegraphics[scale=0.55]{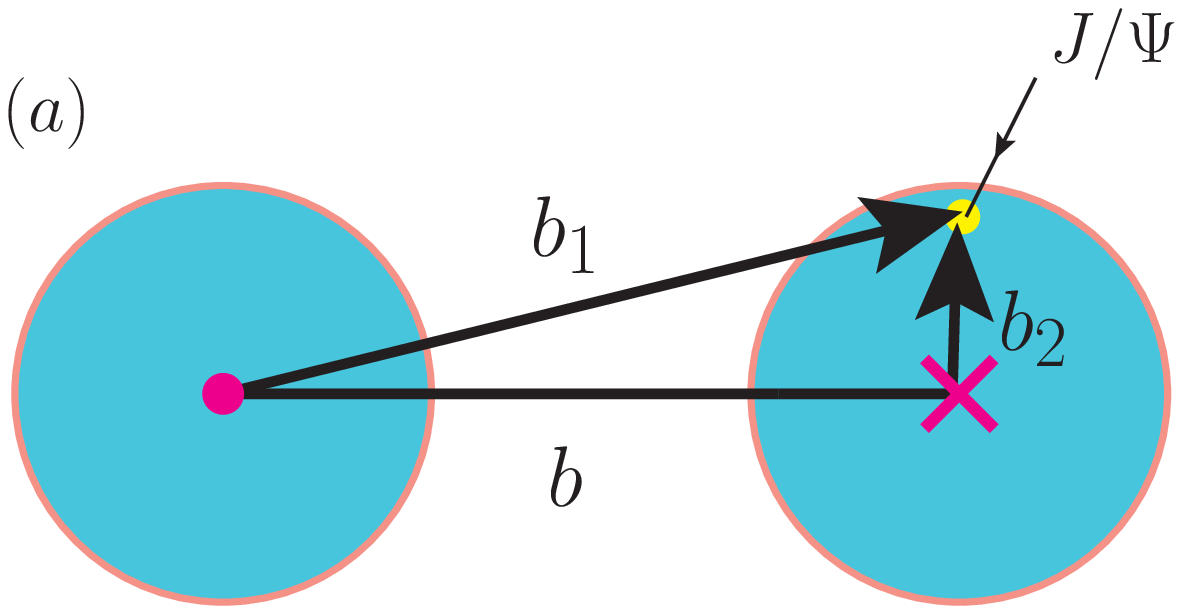}
\includegraphics[scale=0.55]{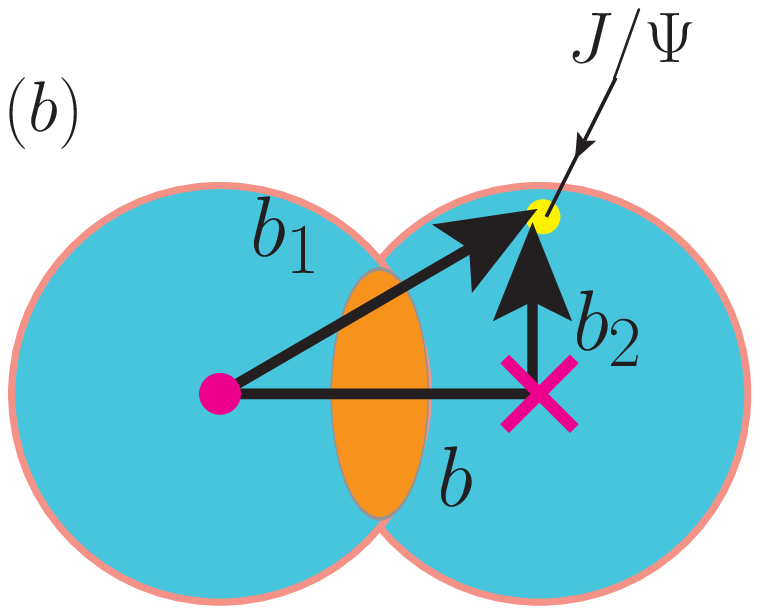}
\caption{Impact parameter picture of the collision and
the production of the $J/\psi$ meson for ultraperipheral (left panel)
and for semi-central (right panel) collisions.
It is assumed here that the first nucleus is the emitter of the photon
which rescatters then in the second nucleus being a rescattering medium.
}
\label{fig:general_situation}
\end{figure}

We shall compare the cross sections for the $J/\psi$ photoproduction 
using the point-like, monopole and realistic form factors
(see Table \ref{table:cross_sections_JPsi}).
But let us have first a closer look at the photon fluxes.
The general situation in the impact parameter space is shown
in Fig.~\ref{fig:general_situation}.
The two-dimensional vectors ${\vec b}$, ${\vec b}_1$ and ${\vec b}_2$
are distances between colliding nuclei, between the photon position
and the middle of the first (emitter) nucleus and between 
the photon position and the middle of the second (medium) nucleus,
respectively. The production of the $J/\psi$ mesons may occur 
provided the photon (hadronic fluctuation) hits the second nucleus,
otherwise the (photo)production is not possible.
The hatched area of overlapping nuclei in the right panel (semi-central collision)
of the figure represents the area in the impact parameter space for which
the situation is not so clear. As will be discussed below we shall
exclude this region to get lower limit for the cross section
for the $A A \to J/\psi$ reaction. In this region of the impact
parameter space the quark-gluon plasma is created and its role
in damping the $J/\psi$ production is not clear.
The lower limit is obtained by assuming the full damping.

\begin{figure}[!h]
\centering
\includegraphics[scale=0.25]{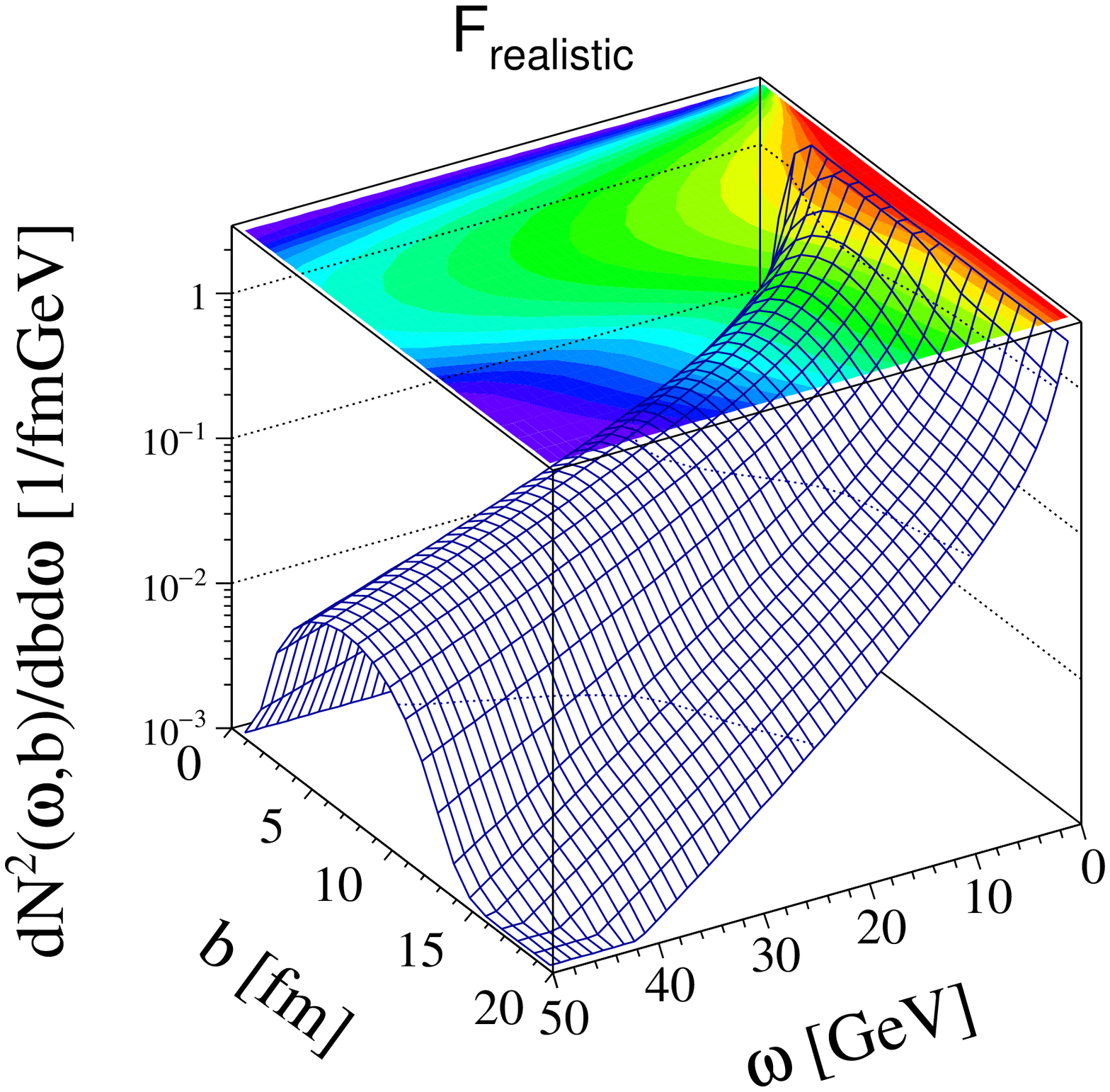}
\includegraphics[scale=0.25]{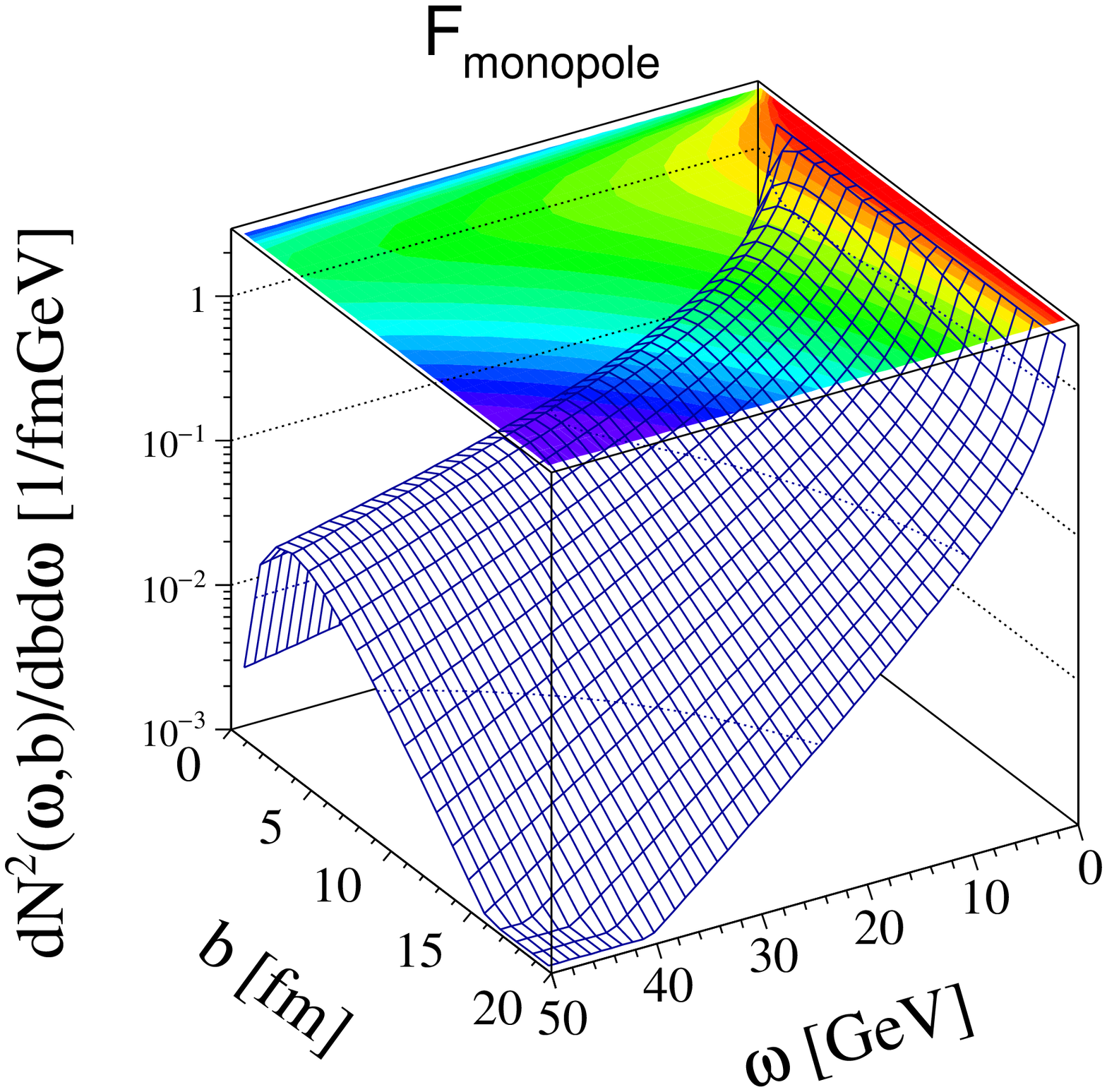}
\includegraphics[scale=0.25]{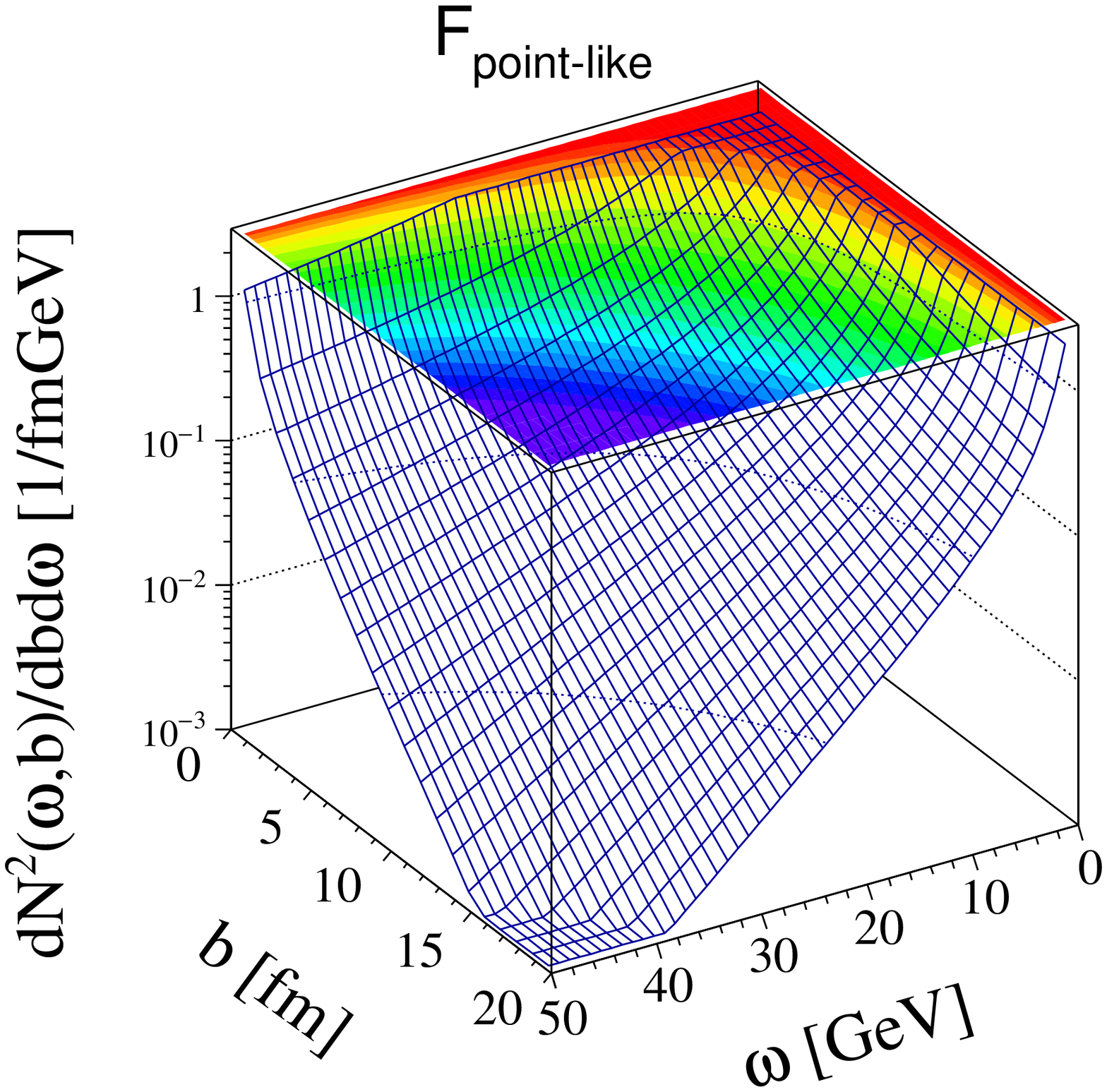}
\caption{Standard photon fluxes calculated for realistic (left panel)
monopole (middle panel) and point-like (right panel) form factors.}
\label{fig:d2N_dbdomega_rmpl}
\end{figure}

Let us start our discussion by showing standard fluxes
used routinely in ultraperipheral collisions.
Fig.~\ref{fig:d2N_dbdomega_rmpl} presents the two-dimensional
photon fluxes (see Eq.~(\ref{eq:N_omegab})) as a function of the distance between 
two colliding nuclei ($b$) and the energy of the photon ($\omega$).
One can observe that the difference between photon fluxes
obtained with realistic (left panel) 
(or monopole (middle panel)) form factors 
and the result for point-like photon source (right panel)
is huge for $b<R_A$ (especially for $b \approx 0$).

\begin{figure}[!h]
\centering
    \includegraphics[scale=0.25]{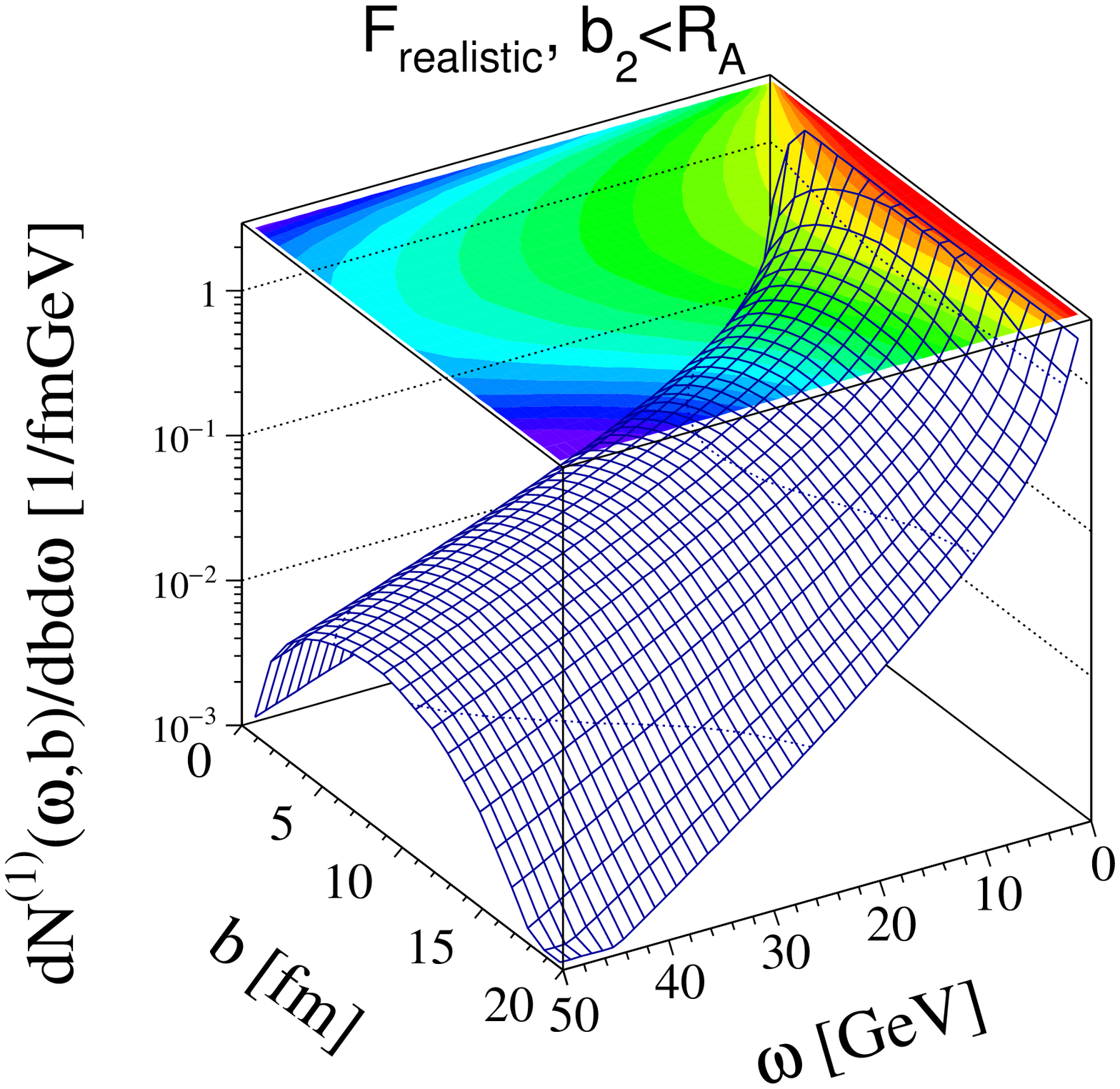}
    \includegraphics[scale=0.25]{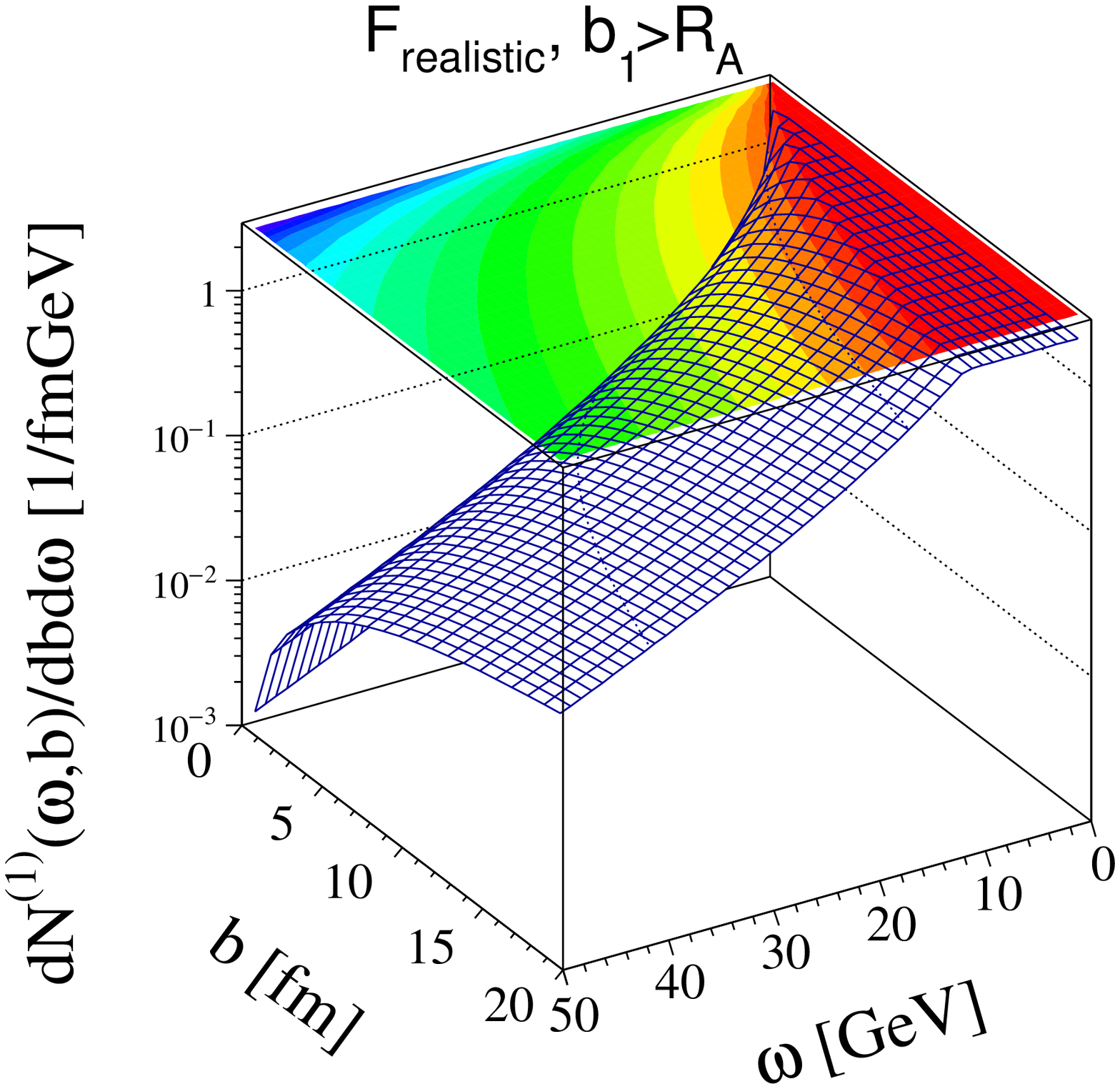}
    \includegraphics[scale=0.25]{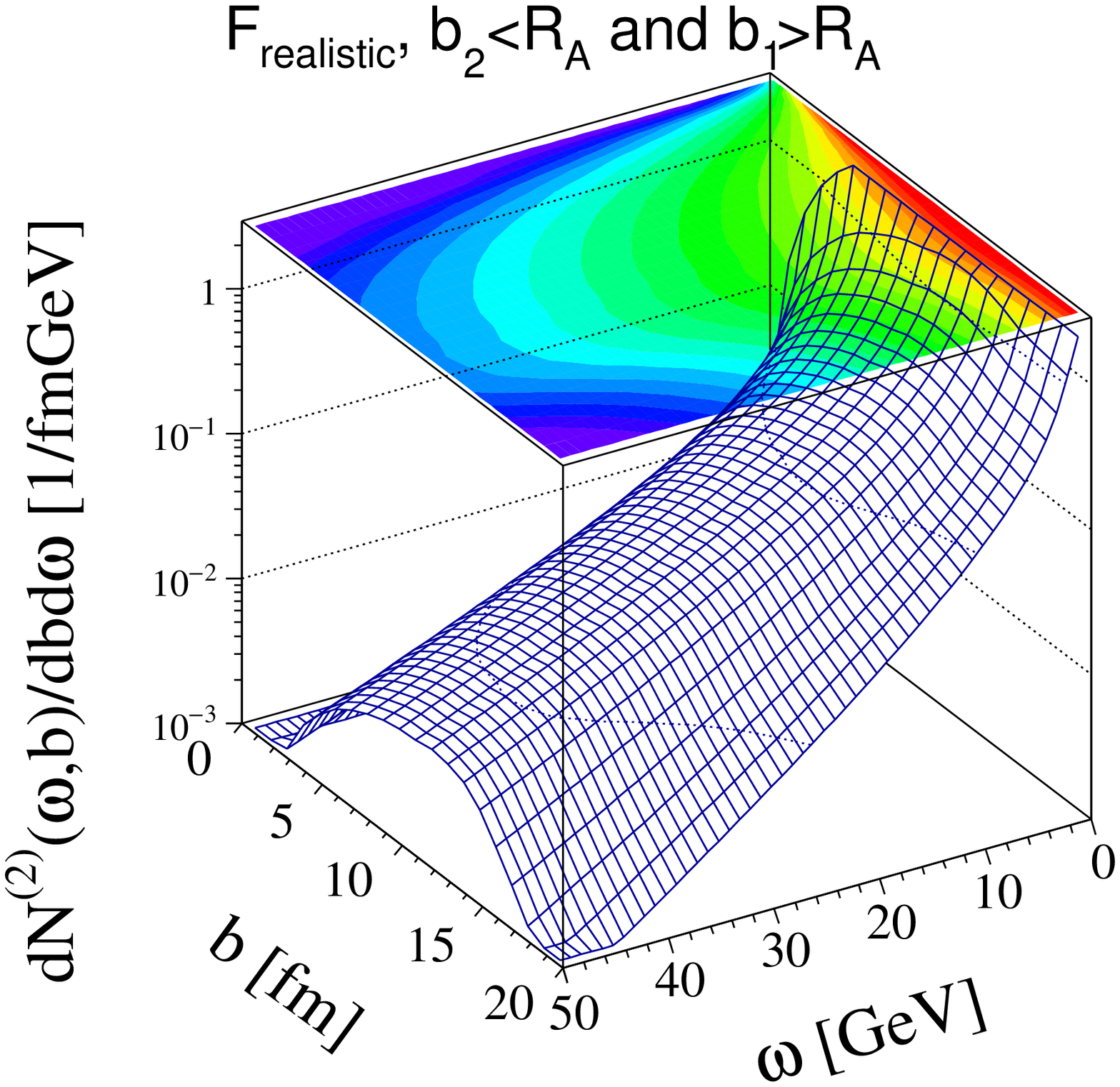}
\caption{Two-dimensional distributions of the photon flux 
in the impact parameter $b$ and in the energy of photon $\omega$
for three different conditions (more in the text).}
\label{fig:d2N_dbdomega}
\end{figure}

Fig.~\ref{fig:d2N_dbdomega} presents the realistic photon flux
which is calculated including extra absorption effects
in three different ways.
The left panel shows the result which is obtained from 
Eq.~\eqref{eq:d2N/domegadb}.
This limit allows a production of vector mesons only inside the second 
nucleus (the one which does not emit the photon).
The second panel is obtained with a condition which follows from the definition
of the absorption factor.  
$\theta( b_1 - R_A)$ in Eq.~\eqref{eq:d2N/domegadb_second}  
allows a production of the meson everywhere except of 
the inside of the tube defined by the emitter.
The last panel shows the result which is obtained from 
Eq.~\eqref{eq:d2N/domegadb_second}.
Here we have included both conditions applied separately  
in the two previous panels of the figure.
This condition can be understood as follows. 
For centrality smaller than $100 \%$
we consider only these cases when particle can be produced
inside the medium excluding this part of the medium which 
coincides with the emitter in the impact parameter space.

\begin{figure}[h]
  \centering
    \includegraphics[scale=0.25]{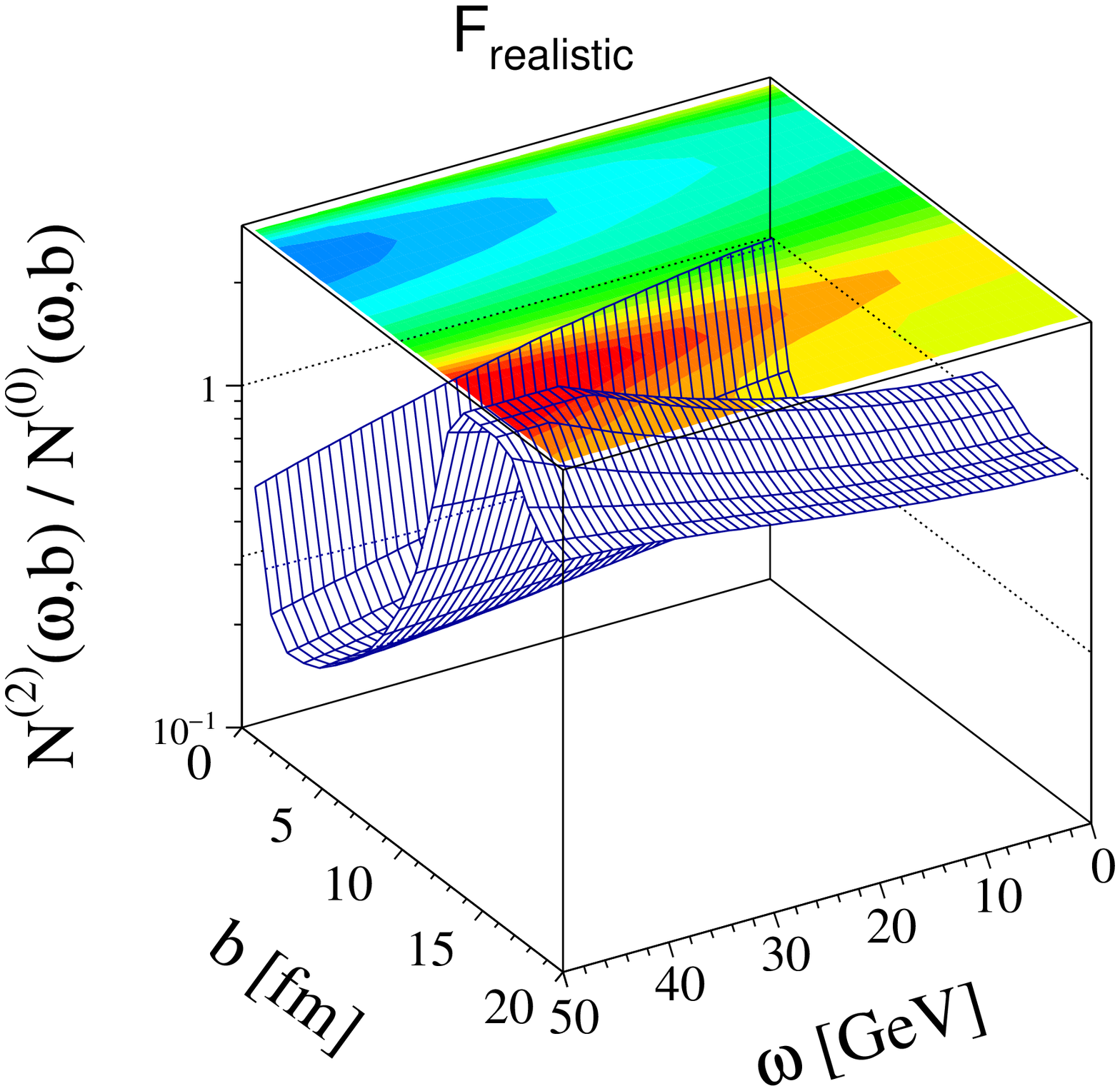}
    \includegraphics[scale=0.25]{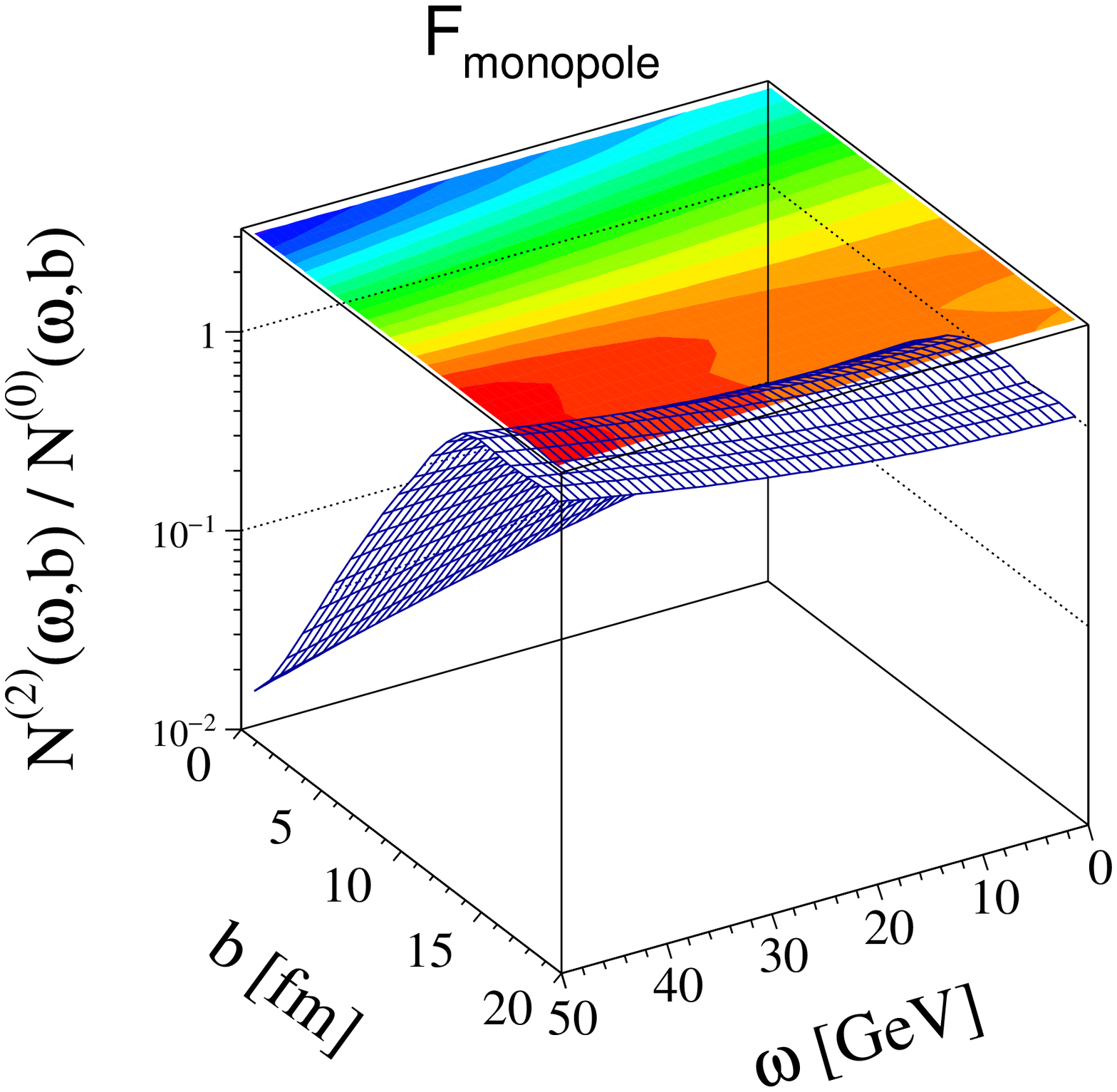}
    \includegraphics[scale=0.25]{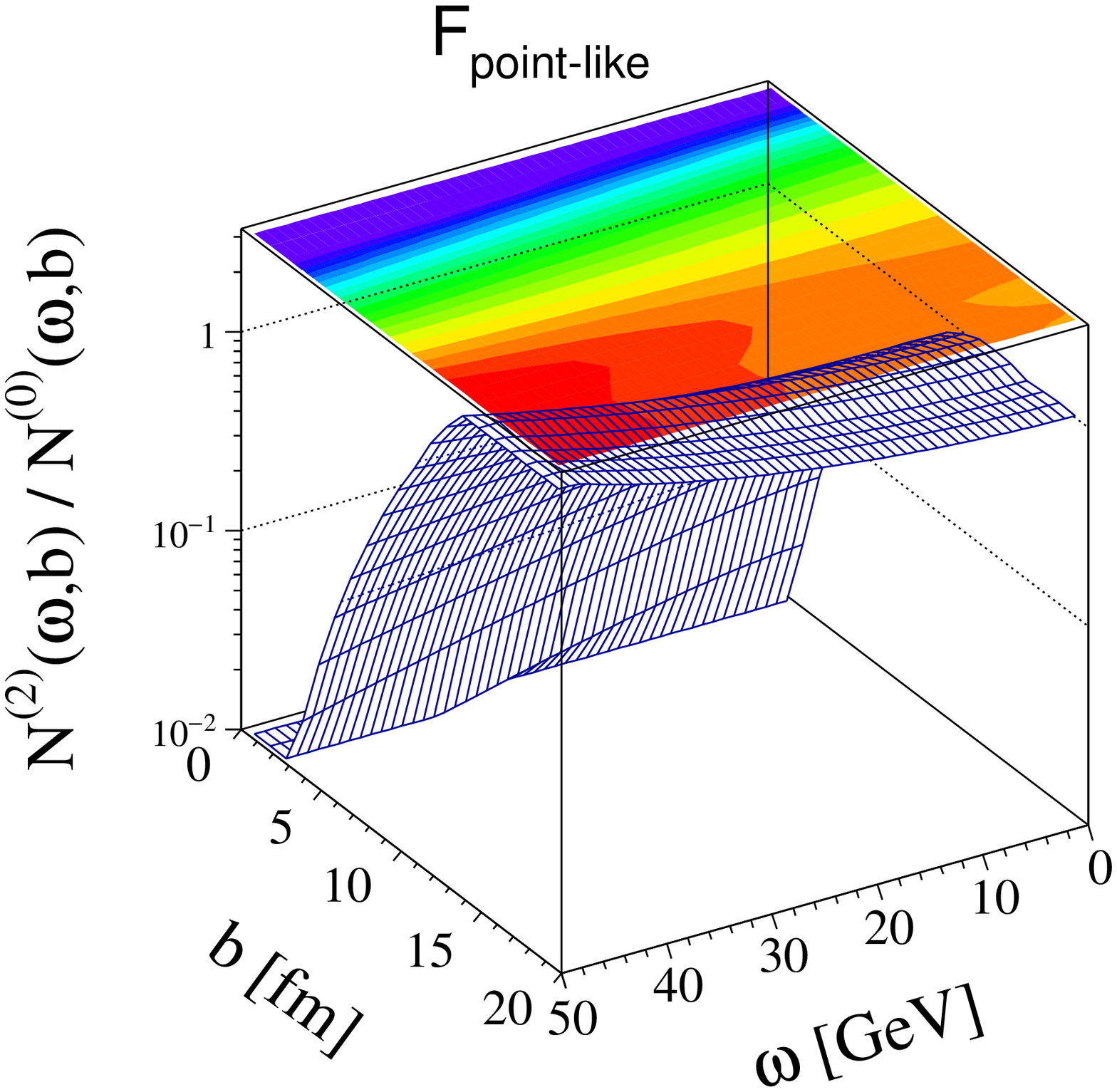}
\caption{The ratio of the differential photon fluxes
in the impact parameter $b$ and energy of the photon $\omega$. 
The left panel shows the case with the realistic form factor,
the middle panel monopole form factor
and the right panel the form factor for point-like charge.}
\label{fig:d2N_dbdomega_ratio}
\end{figure}

We think that it is more pedagogical to inspect the ratio of 
the effective flux in the new approach $N^{(2)}(\omega,b)$
to the standard photon flux (called below $N^{(0)}(\omega,b)$).
Fig.~\ref{fig:d2N_dbdomega_ratio} presents the ratio
of the ''new'' to the standard flux for three types of 
the nuclear form factor.
More precisely, it is a ratio of the result presented in the last
panel of Fig.~\ref{fig:d2N_dbdomega} to the result 
shown in the first panel of Fig.~\ref{fig:d2N_dbdomega_rmpl}.
The three panels represent cases with realistic (left panel),
monopole (middle panel) and point-like (right panel) form factors.
Here we consider lead-lead collisions at the LHC energy 
$\sqrt{s_{NN}}=2.76$ TeV ($\gamma=1471$).
It is shown that the ratio of $N^{(2)}(\omega,b)$ to
$N^{(0)}(\omega,b)$ is almost $1$ in the region of large $b$
for all form factors. However, it is not the case for the semi-central collisions.
These differences very weakly depend on the energy of the emitted photon.
Thus interesting seems to be the cross section 
for different ranges of centrality.

Centrality is a parameter which is often used to characterize the
collision of nuclei at high energies.
To efficiently compare our results with the preliminary ALICE 
data for semi-central collisions \cite{ALICE_presentation} 
we need a simple relation between
impact parameter and the centrality. 
In Ref. \cite{BF_2002} the authors gave an approximate and practical 
geometric relation between them.
In general, centrality of collisions depends on
the impact parameter $b$ and a total inelastic 
nucleus-nucleus cross section
\begin{equation}
c(N) \simeq \frac{\pi b^2\left( N \right)}{\sigma_{inel}} \;.
\end{equation}
In this case the centrality corresponds to a multiplicity higher than $N$.
$b(N)$ expresses the value of the impact parameter for which 
the average multiplicity fulfills the dependence $\overline{n}(b) = N$.
In a purely geometrical picture ($\sigma_{inel} = \pi \left( 2R_A \right)^2$)
we get:
\begin{equation}
c = \frac{b^2}{4R_A^2} \;.
\label{eq:centrality_simply}
\end{equation}
%
\begin{figure}[!h]
\centering
\includegraphics[scale=0.3]{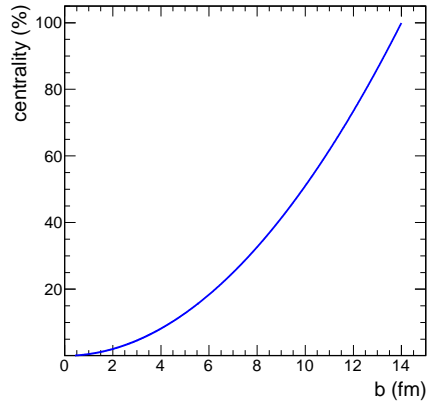}
\caption{Centrality of nuclear collisions 
as a function of the impact parameter as obtained from Eq.~(\ref{eq:centrality_simply}).}
\label{fig:centrality}
\end{figure}

Fig.~\ref{fig:centrality} presents the centrality as a function of
impact parameter. The large value of the centrality
corresponds to the large value of the impact parameter.
The larger value of $b$ the more peripheral collisions.

Let us concentrate now for a while on the second ingredient of the model
(see Eq.~(\ref{eq:probab_P12})) - the $\sigma_{\gamma A \to J/\psi A}$
cross section. 
In our present calculations we use the following sequence of equations:
\begin{equation}
\frac{\mathrm{d}\sigma\left( \gamma p \to J/\psi p; t=0 \right)}{\mathrm{d}t} = 
b_{J/\psi} X_{J/\psi} W_{\gamma p}^{\epsilon_{J/\psi}} \; ,
\end{equation}
\begin{equation}
\frac{\mathrm{d}\sigma\left( J/\psi p \to J/\psi p; t=0 \right)}{\mathrm{d}t} = 
\frac{f^2_{J/\psi}}{4\pi \alpha_{em}} 
\frac{\mathrm{d}\sigma\left( \gamma p \to J/\psi p; t=0
  \right)}{\mathrm{d}t}
\; ,
\end{equation}
\begin{equation}
\sigma_{tot}^2 \left( J/\psi p \right)= 16 \pi
\frac{\mathrm{d}\sigma\left( J/\psi p \to J/\psi p; t=0
  \right)}{\mathrm{d}t}	
\; ,
\end{equation}
\begin{equation}
T_A\left(\textbf{r}\right) = \int \rho \left( \sqrt{r^2+z^2} \right)
  \mathrm{d}z	
\; ,
\end{equation}
%
%
\begin{equation}
\frac{\mathrm{d}\sigma\left( \gamma A \to J/\psi A; t=0 \right)}{\mathrm{d}t}
= \frac{\alpha_{em} \sigma_{tot}^2(J/\psi A)}{4 f^2_{J/\psi}} \; .
\end{equation}
Constants for the $\gamma p \to J/\psi p$ production are obtained
from a fit to HERA data (Ref. \cite{HERA_data}):
$b_{J/\psi}=4$ GeV$^{-2}$, $X_{J/\psi}=0.0015$ $\mu$b,
$\epsilon_{J/\psi}=0.8$ and vector-meson coupling square is
$f^2_{J/\Psi}=4\pi \cdot 10.4$.

The calculation of $\sigma_{tot}(J/\psi A)$
requires incorporating multiple scattering of $J/\psi$ in the nuclear medium.
In the present paper, the calculations are done using both 
classical mechanics ($CM$) and quantum mechanical ($QM$) 
Glauber formula for $\sigma_{tot}(J/\psi A)$ cross section.
''A~quantitative'' and spectacular difference between 
these approaches can be shortly summarized as follows.
In the black disk limit
the classical mechanics approach leads to 
the total cross section equal to $\pi R_A^2$ 
and the quantum mechanical approach implies 
 $\sigma_{tot}(J/\psi A)=2\pi R_A^2$ \cite{FSZ_2003}.
The classical and quantum mechanics expressions for the
$\sigma_{tot}  \left(J/\psi A\right)$ cross section read:
\begin{equation}
\sigma_{tot}^{CM} \left(J/\psi A\right) = \int \mathrm{d}^2 \textbf{r} 
\left( 1-\exp\left( -\sigma_{tot}\left( J/\psi p \right) T_A\left(\textbf{r} \right) \right) \right) \;,
\end{equation} 
\begin{equation}
\sigma_{tot}^{QM} \left(J/\psi A\right) = 2 \int \mathrm{d}^2 \textbf{r} 
\left( 1-\exp\left( -\frac{1}{2} \sigma_{tot}\left( J/\psi p \right) T_A\left(\textbf{r} \right) \right) \right) \;,
\end{equation}
respectively, where $r$ is the distance in the impact parameter space
of the photon (or $c \bar c$ fluctuation) from the middle of 
the nucleus-medium.
These formulae are used to calculate the $\gamma A \to J/\psi A$ cross
section, the main ingredient of our whole approach.

Finally, the total cross section for $\gamma A \to J/\psi A$ reaction
can be written as
\begin{equation}
\sigma_{\gamma A \to J/\psi A} = 
\frac{d\sigma_{\gamma A \to J/\psi A} \left(t=0\right)}{dt}
\int\limits^{t_{max}}_{-\infty} \mathrm{d}t \left| F_A\left(t\right) \right|^2 \;.
\label{eq:sig_tot_gA_JPsiA}
\end{equation}
This cross section is actually a function of energy in the $\gamma A$ system.
The factor $\left| F_A\left(t\right) \right|$ appears here
due to assumed coherent rescattering of the $q\bar{q}$ dipole off a nucleus.
A very good approximation is to use the realistic nuclear charge 
form factor which is defined in Eq.~\eqref{eq:F_real}.
The squared four-momentum transfer: 
$t=-q^2=-(m_{J/\psi}^2/\left(2\omega_{lab}\right))^2$.

\section{Nuclear cross sections for UPC and more central
collisions}

\begin{figure}[!h]
\centering
\includegraphics[scale=0.4]{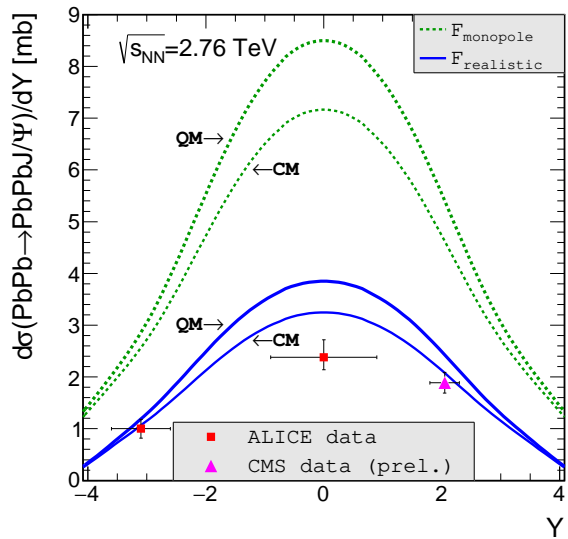}
\caption{Differential cross section for coherent production
of $J/\psi$ meson in UPC as a function of rapidity of the $J/\psi$ meson
compared with the ALICE and CMS data points. We show results for both
realistic and monopole form factor, 
each of them used consistently in Eq.~\eqref{eq:N_omegab}
and \eqref{eq:sig_tot_gA_JPsiA}.}
\label{fig:dsig_dy}
\end{figure}

We start verification of our model for purely ultraperipheral collisions.
Fig.~\ref{fig:dsig_dy} presents results for coherent 
$J/\psi$ photoproduction in the $^{208}$Pb+$^{208}$Pb UPC 
at $\sqrt{s_{NN}}=2.76$ TeV. We show our results for
classical and quantal rescattering in the nucleus-medium
and for different nuclear form factors.
The result with the monopole form factor strongly overestimates
the ALICE and CMS data.
The result with the quantal rescattering is about $15 \%$ larger
than that for the classical rescattering. The difference is much smaller here than
for the photoproduction of $\rho^0$ meson \cite{MKG_thesis}.
The CMS ''data point'' was obtained by correcting a real data
point with at least one neutron in zero-degree calorimeter (ZDC)
using the Monte Carlo program STARLIGHT \cite{STARLIGHT}.
The ALICE data points are taken from Refs. 
\cite{ALICE_2013_y31,ALICE_UPC_midy} 
and the CMS data point is from Ref. \cite{CMS_data}.
Relatively good agreement is obtained for the forward rapidity region
provided the realistic nuclear form factor is used
which supports application of the model also for more central 
($b<R_A+R_B$) collisions especially in the considered rapidity range.

\begin{figure}[!h]
\centering
\includegraphics[scale=0.4]{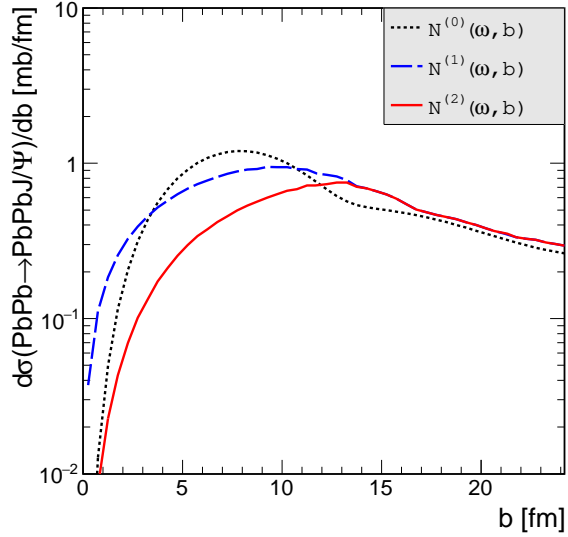}
\caption{Differential cross section for photoproduction
of $J/\psi$ meson as a function of impact parameter 
for $\sqrt{s_{NN}}=2.76$~TeV. 
Different lines correspond to different approximations:
dotted - standard UPC approach, 
dashed - first approximation/correction (upper limit),
solid - second approximation/correction (lower limit). 
Here realistic (charge) nuclear form factor was used.
}
\label{fig:dsig_db}
\end{figure}

In Fig.~\ref{fig:dsig_db} we show the nuclear cross section 
as a function of the impact parameter also for its small values
($b < R_A + R_B$) i.e. for the semi-central collisions.
The different lines correspond to different approximations of photon fluxes
within our approach as described in the figure caption.
The dashed and solid lines represent
upper and lower limit for the cross section.
At larger values of impact parameter $b$ the cross sections obtained with the different
fluxes practically coincide.
At $b < R_A + R_B$ the different approximations give quite different results.
The standard approach used in the literature for UPC
(Eq.~(\ref{eq:N_omegab}))
when naively applied to the semi-central collisions overestimates the cross
section. We will return to this issue in the following.

\begin{table}[h!]
\begin{tabular}{|l|r|r|r|r|r|r|r|}
\hline
Centrality range	 [$\%$]	& 0-10	& 10-30	& 30-50	& 50-70	& 70-90	& 0-100	& UPC \\ \hline
\hline
$\sigma_{tot}^{REAL}$ [mb]						
						& 1.19	& 3.36	& 2.58	& 1.57	& 0.97	& 10.08	& 16.32\\ \hline
$\sigma_{tot}^{MONOP}$ [mb]	
						& 3.64	& 3.99	& 2.16	& 1.36	& 0.99	& 12.58	& 16.26\\ \hline	
$\sigma_{tot}^{PL}$ [mb]
						& 211.00	& 6.11	& 2.46	& 1.50	& 1.06	& 222.68& 16.43\\ \hline
						\hline	
$\mathrm{d}\sigma/\mathrm{d}y_{J/\psi}$ [$\mu$b]	($\Delta y = 1.5$) 
						& $<$318	& $<$290	& 73 
						& 58		& 59			& &  \\
ALICE data				&		&		& $\pm$44$^{+26}_{-27}$	
						&$\pm$16$^{+8}_{-10}$ & $\pm$11$^{+7}_{-10}$ & & \\ \hline	
$\mathrm{d}\sigma^{REAL}/\mathrm{d}y_{J/\psi}$ [$\mu$b]	($\Delta y = 1.5$)
						& 111	& 282	& 196	& 105	& 55		&	&	\\ \hline
$\mathrm{d}\sigma^{MONOP}/\mathrm{d}y_{J/\psi}$ [$\mu$b]	($\Delta y = 1.5$)
						& 398	& 358	& 158	& 86		& 56	 	& 	&  \\ \hline
$\mathrm{d}\sigma^{PL}/\mathrm{d}y_{J/\psi}$ [$\mu$b]	($\Delta y = 1.5$)
						& 9984	& 579	& 185	& 92		& 58		&	& \\ \hline
\end{tabular}
\caption{ \small
Integrated and differential cross section for the photoproduction of $J/\psi$
in $Pb+Pb$ collisions for $\sqrt{s_{NN}}=2.76$~TeV 
calculated with a simple flux factor according 
to Eq.~(\ref{eq:N_omegab})
for the case when realistic (REAL), monopole (MONOP) and point-like (PL) 
form factors are used.
We show results for different ranges of centrality as well as for ultraperipheral events.
The differential cross section is compared with the ALICE experimental
data \cite{ALICE_presentation}.
$\Delta y$ = 1.5 means $y \in$ (2.5,4.0) for brevity.
}
\label{table:cross_sections_JPsi}
\end{table}

In Tab. \ref{table:cross_sections_JPsi} we have collected
the cross sections for six ranges of collision centrality
used in \cite{ALICE_presentation} as well as for UPC. 
We show total (full range of rapidity) and differential 
($\mathrm{d}\sigma/\mathrm{d}y_{J/\psi}$) cross section
for rapidity range of the ALICE experimental studies ($y_{J/\psi}=(2.5-4)$).
Here we take the photon flux given by Eq.~\eqref{eq:N_omegab}.
For each case, the largest cross section appears for centrality range
$10-30 \%$ which corresponds to $b \approx (4.4-7.6)$ fm
according to formula \eqref{eq:centrality_simply}.
One should be aware that the results for the point-like form factor
($F(q)=1$) at very small centrality are not physical.
Here there is a very strong dependence on impact parameter $b$.
The photon flux for very small value of $b$ tends to infinity.
Looking at the agreement of our results with the experimental ALICE data,
one can observe that calculations with realistic form factor
are in fairly good agreement with the experimental data. 
For centrality larger than $30 \%$
the cross section obtained with the monopole form factor is closer
to the experimental value. However, for each form factor
the cross section is larger than that given by the ALICE Collaboration.
Only for centrality range $70-90 \%$ we get a good agreement
with the experimental result. 

It is worth to note here that the nuclear form factor actually appears 
in two formulas in the calculation of the cross section 
for the $A A \to A A J/\psi$ reaction. 
A first one is for the flux of photons (see Eq.~\eqref{eq:N_omegab}) and 
the second one is for the cross section
for the $\gamma A \to J/\psi A$ reaction (see Eq.~\eqref{eq:sig_tot_gA_JPsiA}) 
which is integrated over squared four-momentum transfer.
Only hard sphere form factor was used so far in the literature 
in the context of the $J/\psi$ photoproduction \cite{KN_1999}.
The results presented in Table \ref{table:cross_sections_JPsi}
are calculated for the case when realistic charge distribution 
is included both in Eq.~\eqref{eq:N_omegab} and 
in \eqref{eq:sig_tot_gA_JPsiA},
and for the case when the monopole or point-like form factors are used only in 
calculating the photon flux and realistic form factor is used in Eq. \eqref{eq:sig_tot_gA_JPsiA}.
We suppose that a consistent calculation should use
the same form factor in both places (Eqs. \eqref{eq:N_omegab} 
(or \eqref{eq:d2N/domegadb_second}/\eqref{eq:d2N/domegadb}) and \eqref{eq:sig_tot_gA_JPsiA}).

\begin{table}[h!]
\begin{tabular}{|l|r|r|r|r|r|r|r|}
\hline
Centrality range	 [$\%$]	& 0-10	& 10-30	& 30-50	& 50-70	& 70-90	& 0-100	& UPC \\ \hline
\hline
$\sigma_{tot}^{MONOP}$ [mb]	
						& 8.23	& 9.03	& 4.89	& 3.08	& 2.45	& 28.50	& 38.42\\ \hline	
$\sigma_{tot}^{PL}$ [mb]
						& 17 037.50 & 926.26 & 379.30	& 223.40 & 158.50 & 18795.50 & 2838.30\\ \hline
						\hline	
$\mathrm{d}\sigma/\mathrm{d}y_{J/\psi}$ [$\mu$b]	($\Delta y = 1.5$) 
						& $<$318	& $<$290	& 73 
						& 58		& 59			& &  \\
ALICE data				&		&		& $\pm$44$^{+26}_{-27}$	
						&$\pm$16$^{+8}_{-10}$ & $\pm$11$^{+7}_{-10}$ & & \\ \hline	
$\mathrm{d}\sigma^{MONOP}/\mathrm{d}y_{J/\psi}$ [$\mu$b]	($\Delta y = 1.5$)
						& 892	& 817	& 365	& 201	& 132	 	& 	&  \\ \hline
$\mathrm{d}\sigma^{PL}/\mathrm{d}y_{J/\psi}$ [$\mu$b]	($\Delta y = 1.5$)
						& 1 488 700	& 88 230	& 28 800 & 14 660 & 9 370		&	& \\ \hline
\end{tabular}
\caption{ \small
Integrated and differential cross section for the production of $J/\psi$
in $Pb+Pb$ collisions for $\sqrt{s_{NN}}=2.76$ TeV
for the case when monopole (MONOP) 
and point-like (PL) form factor is included in photon fluxes and 
in the total cross section for the $\gamma A \to J/\psi A$ reaction.
We show results for different ranges of centrality as well as for
ultraperipheral case.
The differential cross section is compared with the ALICE experimental 
data \cite{ALICE_presentation}.
$\Delta y$ = 1.5 means $y \in$ (2.5,4.0) for brevity.
}
\label{table:cross_sections_JPsi_mon_pl}
\end{table}

In Table \ref{table:cross_sections_JPsi_mon_pl}
we have collected the nuclear cross section for
the $PbPb \to Pb Pb J/\psi$ reaction. This table is very similar 
to the previous one, but now we use 
the monopole and point-like form factor
also in the formula which describes the total cross section
for the $\gamma A \to J/\psi A$ reaction \eqref{eq:sig_tot_gA_JPsiA}.
These numbers are much larger than in the previous approach.
We conclude that the point-like approximation
cannot be used to calculate the $\gamma Pb \to J/\psi Pb$ cross section
because it leads to unphysical behaviour for $b<R_A + R_B$.

\begin{table}[h!]
\begin{tabular}{|l|r|r|r|r|r|r|}
\hline
Centrality range	 [$\%$]	& 0-10	& 10-30	& 30-50	& 50-70	& 70-90	& 0-100	 \\ \hline
\hline
$\sigma_{tot}^{REAL}$ [mb]						
						& 1.42	& 2.47	& 2.08	& 1.60	& 1.24	& 9.37	\\ \hline
						\hline	
$\mathrm{d}\sigma/\mathrm{d}y_{J/\psi}$ [$\mu$b]	($\Delta y = 1.5$) 
						& $<$318	& $<$290	& 73 
						& 58		& 59			&   \\
ALICE data				&		&		& $\pm$44$^{+26}_{-27}$	
						&$\pm$16$^{+8}_{-10}$ & $\pm$11$^{+7}_{-10}$ &  \\ \hline	
$\mathrm{d}\sigma^{REAL}/\mathrm{d}y_{J/\psi}$ [$\mu$b]	($\Delta y = 1.5$)
						& 123	& 201	& 160	& 116	& 84	&		\\ \hline
\end{tabular}
\caption{ \small
Integrated and differential cross section for the production of $J/\psi$
in $Pb+Pb$ collisions for $\sqrt{s_{NN}}=2.76$ TeV
calculated with the help of the first approximation 
of the photon flux $N^{(1)}(\omega,b)$ (Eq.~(\ref{eq:d2N/domegadb}))
for the realistic form factor.
The differential cross section is compared with the ALICE 
experimental data \cite{ALICE_presentation}.
$\Delta y$ = 1.5 means $y \in$ (2.5,4.0) for brevity.
}
\label{table:cross_sections_N1}
\end{table}

\begin{table}[h!]
\begin{tabular}{|l|r|r|r|r|r|r|}
\hline
Centrality range	 [$\%$]	& 0-10	& 10-30	& 30-50	& 50-70	& 70-90	& 0-100	 \\ \hline
\hline
$\sigma_{tot}^{REAL}$ [mb]						
						& 0.38	& 1.19	& 1.29	& 1.21	& 1.11	& 5.47	\\ \hline
						\hline	
$\mathrm{d}\sigma/\mathrm{d}y_{J/\psi}$ [$\mu$b]	($\Delta y = 1.5$) 
						& $<$318	& $<$290	& 73 
						& 58		& 59			&   \\
ALICE data				&		&		& $\pm$44$^{+26}_{-27}$	
						&$\pm$16$^{+8}_{-10}$ & $\pm$11$^{+7}_{-10}$ &  \\ \hline	
$\mathrm{d}\sigma^{REAL}/\mathrm{d}y_{J/\psi}$ [$\mu$b]	($\Delta y = 1.5$)
						& 30 	& 88		& 91		& 82		& 72		&		\\ \hline
\end{tabular}
\caption{ \small
Integrated and differential cross section for the production of $J/\psi$
in $Pb+Pb$ collisions for $\sqrt{s_{NN}}=2.76$ TeV
calculated with the help of the second approximation 
of the photon flux $N^{(2)}(\omega,b)$ (Eq.~(\ref{eq:d2N/domegadb_second}))
for the realistic form factor.
The differential cross section is compared with the ALICE experimental 
data \cite{ALICE_presentation}.
$\Delta y$ = 1.5 means $y \in$ (2.5,4.0) for brevity.
}
\label{table:cross_sections_N2}
\end{table}

Table \ref{table:cross_sections_N1} and 
Table \ref{table:cross_sections_N2} collect the cross sections for
different centrality bins for the first (Eq.~(\ref{eq:d2N/domegadb})) and second
(Eq.~(\ref{eq:d2N/domegadb_second})) approximation of the photon flux, respectively.
A fairly good agreement with the ALICE data is obtained 
with the second approximation for all measured centrality bins.

\begin{figure}[!h]
\centering
\includegraphics[scale=0.4]{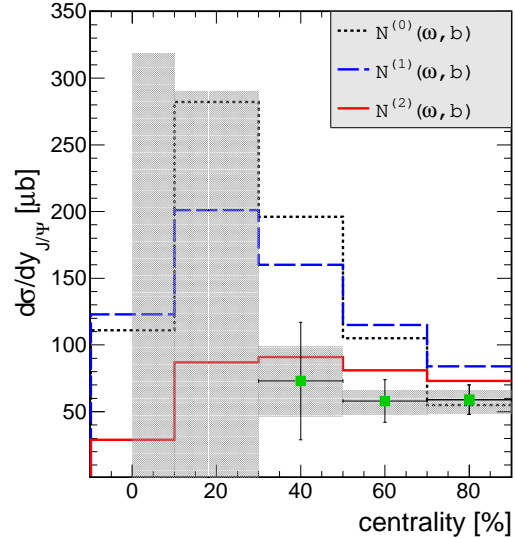}
\caption{$d \sigma / dy$ cross sections for different centrality bins. 
Theoretical results for different
models of the photon flux are compared with the preliminary
ALICE data \cite{ALICE_presentation}.
The shaded area represents the experimental uncertainties.}
\label{fig:flux_hist}
\end{figure}

We summarize the numbers from the tables
in Fig.~\ref{fig:flux_hist}.
We present both statistical and systematic error bars (shaded area).
The ALICE Collaboration could not extract actual values
of the cross section for the two lowest centrality bins. 
The results for standard photon flux exceed the ALICE data.
Rather good agreement with the data is achieved 
for the $N^{(2)}$ photon flux obtained with the realistic
nucleus form factor. The fact that this lower limit
exceeds somewhat the ALICE data may be due to the fact
that the coherent $\gamma A \to J/\psi A$ cross section was used 
in our EPA calculation whereas only spectators may be active
in the production of $J/\Psi$.

\section{Conclusions}

So far photoproduction of vector mesons was considered only 
in the context of ultraperipheral collisions, i.e. in the case
when nuclei survive the collision and appear in the final state,
although the exclusivity is not checked in practice and experimentally 
a limited rapidity gap condition is imposed only.
In this paper we have presented a first theoretical study 
of photoproduction mechanism also in the case when nuclei collide and 
produce quark-gluon plasma and as a consequence considerable number 
of hadrons, mostly pions is produced.
On theoretical side, the nuclear photoproduction in UPC is treated
in the equivalent photon approximation with photon fluxes
and photon-nucleus cross section being the basic ingredients of the
approach.

How to calculate contribution of the photoproduction mechanism
for peripheral and semi-central collisions ($b<R_A+R_B$) 
is not completely clear.
We assume that the whole nucleus produces photons. The photon (or
hadronic photon fluctuation) must
hit the other nucleus to produce the $J/\psi$ meson.
The question arises how to treat the region of overlapping
colliding nuclei in the impact parameter space 
where some absorption may be expected.
We decided to include the effect of the ''absorption'' by modifying
effective photon fluxes in the impact parameter space
by imposing several additional geometrical conditions on
impact parameters (between photon and nuclei and/or 
between colliding nuclei).

In the present paper, as an example,  we have considered a 
vector-dominance based
model which includes multiple scattering effects in addition.
Any other model/approach can be applied in the future.

First we have compared results of our calculations with existing
ALICE and CMS experimental data for UPC.
We have obtained a reasonable agreement with the data especially
in the forward rapidity region when realistic nuclear form factor
is used. 
We have calculated the dependence of the total and differential
cross section on the impact parameter (the distance between
colliding nuclei in the plane perpendicular to the collision axis).
By correcting standard photon fluxes valid only for UPC by 
collision geometry we have calculated cross section
for different centrality bins adequately for the recent ALICE Collaboration
analysis. Our results have been compared with their preliminary data.
We have obtained a reasonable agreement for peripheral and semi-central
collisions and set a lower and upper limits for the cross section for 
the semi-central collisions within our approach.
Our lower limit is, however, model dependent.
Since in our calculations we have used coherent $\gamma A \to J/\psi A$
cross section our lower limit may be overestimated especially
for small impact parameters
where only so-called spectators are active for the $J/\psi$ production.
The time picture of the whole process is not clear to us in the moment.
The rather reasonable agreement of our quite simplified approach 
with the preliminary ALICE data could suggest that the "coherent"
(assumed by the formula used for the $\gamma A \to J/\psi A$ process) 
scattering of the hadronic fluctuation happens before the nucleus 
undergoes the process of deterioration due to
nucleus-nucleus collision and before
the quark-gluon plasma is created. If this is not the case
the "coherent" scattering must be replaced by incoherent one.
Then the transition from UPC to semi-central collisions 
as a function of impact parameter would be more abrupt
then in the present approach.

In the present paper we have performed analysis for a forward
rapidity range. There the $J/\psi$ quarkonia are emitted forward with
large velocity therefore they could potentially escape
from being melted by quark-gluon plasma.
At midrapidities the situation can in principle be different.
It would be therefore valuable if the ALICE Collaboration would repeat 
their analysis also in the midrapidity range
and verify the $p_t \approx$ 0 enhancement.

The photoproduction of $J/\psi$ seems rather special compared to
photoproduction of other objects.
Here the corresponding cross section 
($\frac{\mathrm{d}\sigma_{PbPb \to Pb PbJ/\psi}^{UPC}(y=0)}{\mathrm{d}y} \approx 2$ mb, see Fig. \ref{fig:dsig_dy}) 
is of similar order of magnitude
as the one for $J/\psi$ produced in individual nucleon-nucleon 
collisions or produced in quark-gluon plasma
(simple estimate including 
$\frac{\mathrm{d}\sigma_{pp \to J/\psi}(y=0)}{\mathrm{d}y} \approx 4$ $\mu$b
(Ref. \cite{Zhao_Rapp}),
estimated number of binary collisions and experimental nuclear modification
factor (Ref. \cite{ALICE_midrapidity})
gives $\frac{\mathrm{d}\sigma_{PbPb \to J/\psi}(y=0)}{\mathrm{d}y} \sim 1$ mb
for minimum-bias collisions).
In our simple estimate we get 
$\frac{\mathrm{d}\sigma_{PbPb \to J/\psi}^{photoprod.}(y=0; \,b<R_A+R_B)}{\mathrm{d}y} \sim (0.91 - 1.42)$ mb, where the two numbers are obtained with
$N^{(2)}$ and $N^{(1)}$ fluxes, respectively.
These numbers are of similar size as the contribution of other
mechanisms discussed above.
For (photo)production of other objects the situation may be different.
This was not discussed in the literature so far and requires dedicated studies.

\vspace{1cm}

{\bf Acknowledgments}

We would like to thank Laura Massacrier for a discussion of the results
presented at the EDS Blois 2015 workshop in Borgo (Corsica).
Discussion with Wolfgang Sch\"afer and Ralf Rapp is acknowledged.
This work was partially supported by the Polish grant 
No. DEC-2014/15/B/ST2/02528 (OPUS)
as well as by the Centre for Innovation and Transfer of Natural Sciences
and Engineering Knowledge in Rzesz\'ow.
A~part of the calculations within this analysis was carried out 
with the help of the cloud computer system 
(Cracow Cloud One\footnote{cc1.ifj.edu.pl}) 
of the Institute of Nuclear Physics PAN.



\end{document}